\documentclass[prb,groupedaddress,twocolumn,showkeys,showpacs,floatfix]{revtex4}
\usepackage{color}
\usepackage{amssymb,amsmath}
 \usepackage{graphicx}
\usepackage{epsfig}

\begin{document}

\title{Effect of tilted magnetic field on the magnetosubbands and conductance of bi-layer
quantum wire}
\author{T. Chwiej}
\email{chwiej@fis.agh.edu.pl}
\affiliation{AGH University of Science and Technology, al. A. Mickiewicza 30, 30-059 Cracow,
Poland}
\begin{abstract}
\noindent
The single electron magnetotransport in a vertical bi-layer semiconductor nanowire
made of InAlAs/InGaAs and AlGaAs/GaAs heterostructure is theoretically studied. The magnetic field
is directed perpendicularily to the main (transport) axis of the quantum wire and both non-zero
components of magnetic field, that is the transverse and the vertical ones, allow to change the
magnitude of intra-layer and inter-layer subbands mixing, respectively. We analyze in
detail the changes introduced to energy dispersion relation E(k) by strong titled magnetic field up
to several teslas for a symmetric and an asymmetric confining potential in the growth direction.
These calculated energy dispersion relations are thereafter used to show that the value of
conductance of bi-layer nanowire may jump as well as drop by few conductance quanta when
the Fermi energy is changed what in conjunction with spin Zeeman effect may give a moderately spin
polarized current.
\end{abstract}
\maketitle

\section{Introduction}

The electron transport properties of a nanosystem consisting of two quantum wires being placed
close to each other depends largely on the magnitude of their tunnel
coupling.\cite{eugster,bierwagen,lyo1,mourokh,fischer} For strong and moderate coupling strength
the
electron's wave functions originating from separate wires can hybridize and the magnitude of
hybridization depends naturally on the energy splitting of the electron's subbands and the
electron's wave functions overlap. The last two quantities are particularily sensitive to the
strength of magnetic field and its direction.\cite{lyo3,fischer4} If magnetic field is directed
along the main axis of a coupled wires system, it squeezes the electron's wave functions what leads
to their stronger localization within particular nanowire and in consequence
it weakens the effect of subbands hybridization.\cite{thomas,fischer4}
However, if direction of magnetic field is set perpendicularily to the axis of a nanosystem, then
the Lorentz force pushes the electrons to the edges of the nanowires. The wave functions
being pushed to the central barrier separating the wires can be in such case easily mixed.  Such
inter-wire subbands mixing by perpendicular magnetic field transforms the crossings of hybridized
subbands in the energy dispersion relation E(k) into anticrossings.\cite{shi,lyo1} Inside an
anticrossing a pseudogap is formed what means that the conductance of two coupled nanowires is
lowered
when the Fermi energy level is raised and enters this region. Qualitatively mechanism of the
pseudogaps formation in E(k) spectrum in magnetic field is the same for two laterally or
vertically coupled nanowires systems.
However in practical applications a nanosystem built of two vertically coupled wires has
advantage over a system with two wires aligned laterally since: i) it allows to tune the Fermi
energy in particular wire independently of its value in the second one by means of the top, bottom
and side gates \cite{bielejec,smith,buchholz,hew,fischer} and, ii) each of the three components of
magnetic field modifies in a different manner the conductance of a
nanosystem\cite{fischer2,mourokh} giving thus an opportunity to control over a single
subband.\cite{fischer2}

In this paper, we theoretically analyze the effect of tilted magnetic field on a single electron
transport in a bi-layer quantum wire made of a double inverted heterostructure like
InAlAs/InGaAs and AlGaAs/GaAs. We assume the electron current flows in the wire (along x axis)
without scattering i.e. the electron transport is ballistic, layers are vertically aligned in z
(growth)
direction one over another and the surrounding confining potential is formed by the rectangular
barriers. Direction of magnetic field is perpendicular to the wire axis, that is  only its
vertical
($B_{z}$) and transverse ($B_{y}$) components have non-zero values and both can be changed
independently with precision.\cite{tlitedB} Such approach give us an ability to tune an interlayer
and intralayer modes coupling by changing $B_{y}$ and $B_{z}$, respectively.
First, we discuss how the simultaneous mixing of the vertical and the transverse eigenstates
for the assumed double-well confining potential modifies the energy dispersion relation $E(k)$ of
bi-layer nanowire. Next the oscillating behaviour of magnetosubbands in vicinity of $k=0$ is
considered in context of the conductance variations as function of Fermi energy.
We show that for a nanosystem working in a ballistic regime these oscillations give contribution to
the wire conductance which may jump as well as drop by few conductance quanta when Fermi energy
level is successively raised or lowered between two neighbouring pseudogaps.
In the last part of this paper we discuss a possibility of application of  bilayer quantum wire  as
a source of partly spin polarized current for moderate Fermi energies.

Paper is organized as follows. In Sec.\ref{Sec:teo} we present theoretical model used in
calculations,
properties of magnetosubbands for tilted magnetic field are discussed
in Sec.\ref{Sec:res} while in Sec.\ref{Sec:pol} we present the potential application of
 pseudogaps appearing in energy spectrum for partial spin polarization of conductance. We
end up with conclusions given in Sec.\ref{Sec:con}.

\section{Theoretical model}
\label{Sec:teo}

The confining potential in a conventional semiconductor quantum wire can be formed electrostatically
by gating the 2DEG\cite{hew}, etching of
nano-grooves\cite{fischer} on the layered nanostructure that holds 2DEG few tens of nanometers
beneath the surface or by cleaved-edge overgrowth.\cite{huber} In all cases the 2DEG is formed
within a square well created by double inverted heterojunction.\cite{fischer} The first three
methods give soft lateral confining potential
which was widely used in theoretical works before\cite{window-the,ihnatsenka} while the last one
generates rectangular confinement that we have adopted for this work.
We consider a quantum wire in which the electrons can move freely along x axis but
their motion in y-z plane is quantized due to the rectangle shapes of external barriers.
The quantum well is defined for $y\in[-a,a]$ and $z\in[-b,b]$ with high confining potential
outside this region. Throughout this paper we use $a=50\textrm{ nm}$ and $b=15\textrm{ nm}$.
For simplicity  we assume the barrier surrounding the wire is infinite
while the confining potential inside the channel depends only on position in the growth direction
(z-axis) i.e. $V(\pmb{r})=V(z)$. We model the confining potential by formula:
\begin{equation}
 V(z)=V_{max}[sin\left((1+z/b)\pi/2 \right) +\alpha sin\left(\pi(1+z/b)\right)]
\label{Eq:vz} 
\end{equation}
which describes the potential with maximum localized in a central region of original well
[non-zero parameter $\alpha$ breaks the symmetry in $V(z)$].
Such potential is formed within a wide quantum well when one delta doping layer is placed
below  and another one above the well. Then, the positively ionized dopants effectively lower the
confining potential near both edges of the well giving thus a bi-layer coupled system within a
single nanowire.\cite{fischer4,delta2}
Depth of an upper well, or in other words the value of parameter $\alpha$, can be adjusted by
changing the voltage applied to the central top gate\cite{thomas} which may cover the whole
structure\cite{distortion}  or to the top split gate.\cite{fischer4}

\begin{figure}[htbp!]
\hbox{
	\epsfxsize=80mm
        \epsfbox[50 370 780 743] {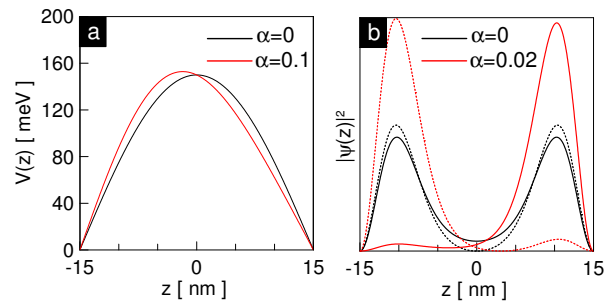}
        \hfill}
\caption{(Color online) a) Cross-section of a confining potential in z direction for
$\alpha=0$ and $\alpha=0.1$ (left and right axes are the infinite barriers). b) Probability density
distributions of two lowest eigenstates of vertical quantization for symmetric
($\alpha=0$ - black color) and nonsymmetric ($\alpha=0.02$ - red
color) confining potentials. Continous curves stand for the ground state while dashed lines for the
first excited one. In (a) and (b) $V_{max}=150\textrm{ meV}$.}
\label{Fig:vz}
\end{figure}

An example of a such confining potential is showed in Fig.\ref{Fig:vz}(a).
A bi-layer system can be also formed by  stacking two quantum wires one above the other during the
epitaxial growth with very narrow tunnel barrier separating them.\cite{fischer,fischer2,buchholz}
Since all effects, we investigate here, depend mainly on the energy difference between two 
lowest eigenstates of vertical quantization, an actual shape of vertical confinement is of little
importance and the results presented below are representative for both types of
confinements. Both layers, the upper and the lower one are pierced by magnetic field which has
$B_{x}=0$ whereas the values of two other components
can be freely changed. Since magnetic field lifts spin degeneracy,  we limit our
considerations to subbands with spins set parallel to magnetic field until otherwise stated.
The energies for subbands with antiparallel spins can be simply obtaned by adding spin-Zeeman
splitting energy $\Delta E_{Z}=g\mu_{B}B$.
In calculations we use a non-symmetric  vector potential
$\pmb{A}=[zB_{y}-yB_{z},0,0]$ for which the single electron Hamiltonian reads:
\begin{equation}
 H=-\frac{\hbar^2}{2m^{*}}\nabla^2+\frac{Iq\hbar}{m^{*}}A_{x}\frac{\partial}{\partial x}
 +\frac{q^2}{2m^{*}}A_{x}^{2}
 \label{Eq:ham0}
\end{equation}
where: $I$ is an imagnary unit, $q=-e$ is an electron's charge and $m^{*}$ is its effective
mass ($m^{*}=0.067$ for GaAs and $m^{*}=0.04$ for InGaAs).
The eigenstates of Hamiltonian given by Eq. \ref{Eq:ham0}) can be expressed as the linear
combination of products of the plane waves for x direction and the eigenstates for the transverse
(y-axis) and vertical (z-axis) directions.
We define the electron's wave function for p-th subband and the wave vector k as follows:
\begin{equation}
 \Psi_{p,k}(\pmb{r})=e^{Ikx} \sum_{m=1}^{M}\sum_{i=1}^{N}c^{(p)}_{m,i}\cdot f_{m}(z)\cdot
\varphi_{i}(y)
\label{Eq:psipk}
\end{equation}
In Eq. \ref{Eq:psipk}, $f_{m}(z)$ and $\varphi_{i}(y)$ are  the basis functions for quantization in
z and y directions, respectively. Due to infinite barriers surrounding the quantum wire, the
normalized basis functions for y direction are simply:
\begin{equation}
 \varphi_{i}(y)=sin\left[i\cdot\pi(1+y/a)/2\right]/\sqrt{a},\quad i=1,2,\ldots,N
 \label{Eq:fiy}
\end{equation}
The basis functions $f_{m}(z)$ have been found by solving the
eigenproblem for z direction i.e. $\widehat{h}_{z}f_{m}=E^{z}_{m}f_{m}$  with Hamiltonian
$h_{z}=-(\hbar^2/2/m^{*})\partial^2/\partial z^2+V(z)$. 
The coefficients of linear combination $c^{(p)}_{m,i}$ appearing in
Eq. \ref{Eq:psipk}
are real numbers which are strictly 0 or 1 when magnetic mixing is absent
($B_{y}=B_{z}=0$) otherwise they fulfill condition $|c^{(p)}_{m,i}|\le 1$.

\begin{figure}[htbp!]
\hbox{
	\epsfxsize=80mm
        \epsfbox[60 320 540 550] {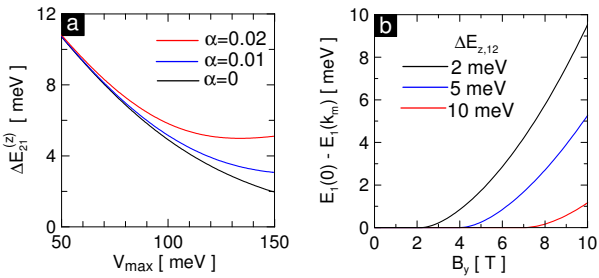}
        \hfill}
\caption{(Color online) a) Energy difference  $\Delta E_{21}^{(z)}$ between the first and second
eigenstates for vertical direction in dependence on $V_{max}$ and for different values of
$\alpha$.
b) An energy difference between maximum ($k=0$) and minimum ($k_{m}$) of
lowest subband in dependence on $B_{y}$. In (b) results were obtained for
 $V_{max}=150\ \textrm{meV}$ and $\alpha=0$.}
\label{Fig:de21}
\end{figure}

The maximum in the confining potential divides the original well into two narrower tunnel-coupled
wells
which for $\alpha=0$ have the same eigenenergies.
Figure \ref{Fig:vz}(b) shows the probability densities for the ground [$f_{1}(z)$] and the first
excited [$f_{2}(z)$] eigenstates of Hamiltonian $h_{z}$.
We notice that for large amplitude of $V(z)$ [see Eq.\ref{Eq:vz}], the
densities are localized mainly in two
narrower wells but mutually overlap to some extent. Since energy difference for these
eigenstates  [$\Delta E^{(z)}_{21}=E^{(z)}_{2}-E^{(z)}_{1}$] is small, even a small
distortion in the confinement [$\alpha=0.02$ in
Figs.\ref{Fig:vz}(b)] can significantly mix them what may result in their spatial separation [red
curves
in Fig.\ref{Fig:vz}(b)]. Therefore, these two thin wells form two transport layers and the nanowire
becomes in fact a bi-layer system. To get deeper insight into nature of this process, we
calculate $\Delta E^{(z)}_{21}$ as function of $V_{max}$. This dependency is showed in
Fig.\ref{Fig:de21}(a). In spite of $\alpha's$ value, $\Delta E^{(z)}_{21}$ decreases if $V_{max}$
grows but the lowest $\Delta E^{(z)}_{21}$ we get always for $\alpha=0$. For $\alpha>0$,
the ground state is localized in the upper well what minimizes its energy  while the
first excited one occupies the lower well because of orthogonality constrictions but for the price
of its increased energy.\cite{distortion}
Due to a large energy separation of higher 
eigenstates $f_{m}(z)$ ($m=3,4,\ldots$) we have neglected them in calculations and use $M=2$ in
Eq.\ref{Eq:psipk} while the number of transverse modes $\varphi_{i}(y)$ was limited to $N=30$.

Having the wave functions defined by Eq.\ref{Eq:psipk}, we may used them to transform
an original Hamiltonian given by Eq.\ref{Eq:ham0} to much simpler algebraic form.
For this puropse, we first eliminate x variable from Hamiltonian
$H$ by averaging it over this variable:
$H_{y,z}=\langle e^{-Ikx}|H|e^{Ikx}\rangle$.
Reduced Hamiltonian then reads:
\begin{equation}
 H_{yz}=-\frac{\hbar^2}{2m^{*}}\left(\nabla^2_{y,z}+k^2 \right)-\frac{q\hbar k}{m^{*}}A_{x}
 +\frac{q^2}{2m^{*}}A_{x}^{2}+V(z)
\end{equation}
Next, we calculate $H_{z}=\langle
\varphi_{i}(y)|H_{yz}|\varphi_{j}(y) \rangle$ and thereafter $H^{eff}=\langle f_{l}(z)
|H_{z}|f_{m}(z) \rangle$.  The effective Hamiltonian has the following form:

\begin{equation}
H^{eff}=H^{eff}_{0}+H^{eff}_{1}+H^{eff}_{2}+H^{eff}_{3}
\label{Eq:hef}
\end{equation}

\begin{eqnarray}
 H^{eff}_{0}&=&\left( \frac{\hbar^2
\gamma_{j}^{2}}{2m^{*}}\delta_{i,j}+E_{m}^{z}\delta_{i,j}+\frac{1}{2m^{*}}Y^{(2)}_{i,j}
\right)\delta_{l,m}
\label{Eq:hef0}\\
H^{eff}_{1}&=&\frac{m^{*}\omega^{2}_{y}}{2}Z_{l,m}^{(2)}\delta_{i,j}
\label{Eq:hef1}\\
H^{eff}_{2}&=&-\omega_{y}\hbar k \delta_{i,j}Z_{l,m}^{(1)}
\label{Eq:hef2}\\
H^{eff}_{3}&=&-m^{*}\omega_{y}\omega_{z}Y^{(1)}_{i,j}Z_{l,m}^{(1)}
\label{Eq:hef3}
\end{eqnarray}
In above array of equations we have assumed:
$\gamma_{j}=j\pi/2/a,\ j=1,2,\ldots,N$;
$\omega_{y}=qB_{y}/m^{*}$, $\omega_{z}=qB_{z}/m^{*}$; 
$Z_{l,m}^{(1)}=\langle f_{l}|z|f_{m}\rangle$;
$Z_{l,m}^{(2)}=\langle f_{l}|z^2|f_{m}\rangle$;
$Y_{i,j}^{(1)}=\langle \varphi_{i}|y|\varphi_{j}\rangle$;
$Y_{i,j}^{(2)}=\langle \varphi_{i}|(m^{*}\omega_{z}y+\hbar k)^2|\varphi_{j}\rangle$;
and $\delta_{i,j}$ is a Kronecker's delta. Indices $(i,j)$ stand for transverse modes
(defined in Eq.\ref{Eq:fiy}) while pair of  $(l,m)$ mark the vertical ones.
The effective Hamiltonian has matrix form which for only two-element basis
$\{f_{m}(z)\}$ becomes $2\times2$ block matrix with real elements:
\begin{equation}
\pmb{H}^{eff}=
\left[
\begin{array}{cc}
\pmb{H_{11}}&  \pmb{H_{12}}\\
\pmb{H_{21}}&  \pmb{H_{22}}
\end{array}
\right]
\label{Eq:hefb}
\end{equation}
Since all  terms appearing in Eqs.\ref{Eq:hef0}-\ref{Eq:hef3} are real,
we immediately get condition $H_{21}=H_{12}$.
The only term depending on $\delta_{l,m}$ is $H_{0}^{eff}$ and therefore it contributes to
diagonal submatrices $H_{11}$  and $H_{22}$ (their rank equals N that is the number of basis states
$\varphi_{i}$). In consequence  $H_{0}^{eff}$ does not mix z-eigenstates.
If confining potential $V_{z}$ is symmetric ($\alpha=0$)  then the
wave function $f_{1}(z)$ has
even parity (ground or bounding state) while $f_{2}(z)$ has odd parity (first excited or antibonding
state). Since the term $Z_{l,m}^{(2)}$ has itself even parity,
the matrix elements $H_{1}^{eff}$ give contributions only to diagonal submatrices, otherwise
($\alpha>0$) they contribute also to off-diagonal submatrix $H_{12}$. On the other hand, term  
$Z_{l,m}^{(1)}$ has odd
parity for $\alpha=0$. It mixes vertical states $f_{1}$ and $f_{2}$ and in consequence gives
contribution to $H_{12}$ due to $H_{2}^{eff}$ and $H_{3}^{eff}$ terms. Notice,
that for $\alpha>0$ these Hamiltonians give contributions to diagonal matrices $H_{11}$ and
$H_{22}$.

\section{Results}
\label{Sec:res}

\subsection{Two-state model for $B_{z}=0$}
\label{Sec:bz0}
\begin{figure}[htbp!]
\hbox{
	\epsfxsize=80mm
        \epsfbox[80 285 515 540] {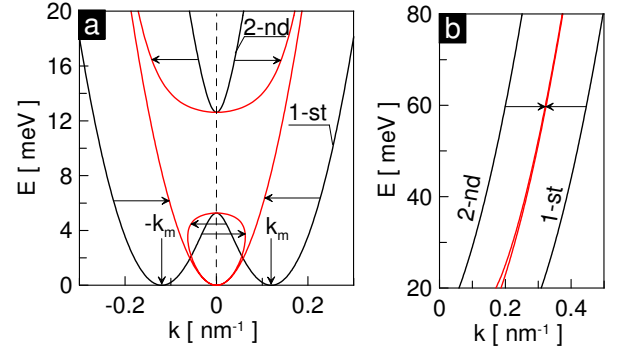}
        \hfill}
\caption{(Color online) Energy dispersion relation in function of canonical wave vector k (black
line) and kinetic wave vector $k_{kin}$ (red line) for two lowest subbands. Figure (b)
is continuation of (a) for higher energy and shows degeneracy of both subbands as function of
kinetic wave vector $k_{kin}$. Horizontal arrows on (a) show directions of wave vector
transformation $k\to k_{kin}$ while  vertical ones mark energy minima in first subband. Parameters
used in calculations: $B_{y}=10\ \textrm{T}$, $B_{z}=0$, $\alpha=0$, $\Delta E^{(z)}_{21}=5\
\textrm{meV}$. All energies are given with respect to a bottom of the lowest subband.}
\label{Fig:e1e2}
\end{figure}

For $B_{z}=0$ there is no mixing between basis states $\{\varphi_{i}(y)\}$ since the term
$Y^{(1)}_{i,j}$  disappears in $H^{eff}$ while term $Y^{(2)}_{i,j}$
reduces to $Y^{(2)}_{i,j}=\hbar^{2}k^{2}\delta_{i,j}$ due to $\omega_{z}=0$ [see
Eqs.\ref{Eq:hef0}-\ref{Eq:hef3}].
This gives us possibility to limit our considerations for a moment to the case with $N=1$
keeping in mind that energy dispersion $E(k)$ for transverse modes $\varphi_{i}$ with
indcies $i>2$ are strictly replicas of that with  $i=1$. Then, only diagonal elements of  $H^{eff}$
are shifted towards higher energy in the same manner. The two-state Hamiltonian now reads:

\begin{widetext}
\begin{eqnarray}
 \pmb{H}^{eff}&=&
 \left[
\begin{array}{cc}
H_{0}^{eff}+H_{1}^{eff}&H_{2}^{eff} \\
H_{2}^{eff}&H_{0}^{eff}+H_{1}^{eff} \\
\end{array}
 \right]\\
 &=&
 \left[
\begin{array}{cc}
\frac{\hbar^2(\gamma_{1}^2+k^2)}{2m^{*}}+E_{1}^{(z)}+\frac{m^{*}\omega_{y}^2}{2}Z_{1,1}^{(2)}&
-\omega_{y}\hbar k Z_{1,2}^{(2)}\\
-\omega_{y}\hbar k Z_{1,2}^{(2)}&
\frac{\hbar^2(\gamma_{1}^2+k^2)}{2m^{*}}+E_{2}^{(z)}+\frac{m^{*}\omega_{y}^2}{2}Z_{2,2}^{(2)}\\
\end{array}
 \right]
 \label{Eq:hi}
\end{eqnarray}
\end{widetext}
Eigenvalues of $\pmb{H}^{eff}$ can be written as:
\begin{equation}
 E_{1,2}=\frac{\hbar^{2}k^2}{2m^{*}}+A_{3}\pm \frac{|A_{1}|}{2}
 \sqrt{1+\left(\frac{A_{2}}{A_{1}}\right)^{2}k^2}
 \label{Eq:e21}
\end{equation}
where we have used following abbrievations:
$A_{1}=\left(E_{2}^{(z)}-E_{1}^{(z)}\right)+\frac{m^{*}\omega_{y}^{2}}{2}
\left(Z_{2,2}^{(2)}-Z_{1,1}^{(2)} \right)$;
$A_{2}=2\omega_{y}\hbar Z_{12}^{(1)}$; 
$A_{3}=\frac{\hbar^2
\gamma_{i}^2}{2m^{*}}+\left(E_{1}^{(z)}+E_{2}^{(z)}\right)/2
+\frac{m^{*}\omega_{y}^2}{4}\left(Z_{1,1}^{(2) } +Z_ { 2 , 2 }^{(2)}\right)$.
Energy dispersion relation $E(k)$ for these two subbands are displayed in
Fig.\ref{Fig:e1e2}(a) (black color). In first subband there are three extremums: two
minimums separated by maximum localized at $k=0$.
Localization of minimums can be found by imposing condition on dispersion relation $\partial
E/\partial k|_{k=k_{m}}=0$ what gives:
\begin{equation}
k_{m}=\pm\sqrt{\left(\frac{m^{*}}{2\hbar^2}\right)^2 A_{2}^{2}-\left(\frac{A_{1}}{A_{2}}
\right)^2}
\label{Eq:km}
\end{equation}
Value of $k_{m}$ depends on $A_{1}$ and $A_{2}$ and therefore, for a fixed geometry  and
values of material parameters of nanowire, the non-zero value of $k_{m}$ depends only on $B_{y}$ due
to
requirement of non-negativity of  expression in square root in Eq.\ref{Eq:km}. Minimal value
of $B_{y}$ that gives $k_{m}>0$ can be estimated from the following formula:
\begin{equation}
 B_{y}^{min}>\sqrt{
\frac{2m^{*}}{q^2}
 \frac{\left(E_{2}^{(z)}-E_{1}^{(z)}\right)}
 {4 \left(Z_{1,2}^{(1)}\right)^2-\left(Z_{2,2}^{(2)}-Z_{1,1}^{(2)}\right)}}
 \label{Eq:bmin}
\end{equation}
For $\alpha=0$, when energy difference $\Delta E_{21}^{(z)}$ is
determinated by value of $V_{max}$, the increase of $V_{max}$  decreases value of $\Delta
E_{21}^{(z)}$ and consequently smaller $B_{y}^{min}$ is needed for lateral minima to appear. Such
dependence is displayed in
Fig.\ref{Fig:de21}(b) which shows a difference between  maximum and minimum of energy for
first subband in function of $B_{y}$ value. Simply, the lower $\Delta E_{21}^{(z)}$ is, the lower
$B_{y}$ is needed for this difference to have non-zero value. 
In close vicinity of $k=0$ the squared term in Eq.\ref{Eq:km} can be expanded in  power  series of
k. Neglecting the terms with exponents greater than 3 which are small, we get
the parabolic shape of energy dispersion:
\begin{equation}
 E_{1,2}\left|_{k\to 0}\right.=A_{3}\pm \frac{|A_{1}|}{2}+k^{2}\left( \frac{\hbar^2}{2m^{*}}
\pm \frac{1}{4}\frac{A_{2}^{2}}{|A_{1}|}\right)
\label{Eq:k0}
\end{equation}
where sign $+(-)$ corresponds to upper (lower) subband.
For $k=0$, the energy
difference between both subbands equals $|A_{1}|$ and depends on the sum of two differences: i)
$\left(E_{2}^{(z)}-E_{1}^{(z)}\right)$ and, ii) $\left(Z_{2,2}^{(2)}-Z_{1,1}^{(2)}\right)$.
Let us notice that, for increasing value of $V_{max}$ both 
basis states for vertical direction i.e. $f_{1}(z)$  and $f_{2}(z)$ becomes degenerated
[Fig.\ref{Fig:de21}(a)] with
similar densities [Fig.\ref{Fig:vz}(b)]. For this reason, the energy difference between two lowest
subbands  gets smaller when $V_{max}$ is increased for $k=0$.
Appearance of two additional energy minimums in magnetosubbands brings severe
consequences for electron transport. If $B_{y}$ is large enough to create energy minima then the
kinetic wave vector $k_{kin}=\langle k+qA_{x}/\hbar\rangle$ becomes negative
for  $k\in(0,k_{m})$. Similarly, due to symmetry of a confinig potential, for negative k in the
range $k\in(-k_{m},0)$, the effective wave vector becomes positive. Any oscillations in energy
depending on canonical wave vector k produce thus negative energy dispersion relations.
To study this problem in detail we have plotted the energies of two lowest
subbands in function of $k_{kin}$ in Fig.\ref{Fig:e1e2} (red color).
After transformation $k\to k_{kin}$, the dependence of electron's energy on $k_{kin}$  becomes
ambiguous for the first subband when $k_{kin}$ is small. It consists of two curves:
the closed loop surrounded by a parabolic branch. The horizontal arrows in Fig.\ref{Fig:e1e2}
indicates directions of the wave vector's transformations.
First, let us notice that for the closed loop, $k\in[-k_{m},0]$
transforms to $k_{kin}>0$ and due to symmetry $k\in[0,k_{m}]$ transforms to $k_{kin}<0$.
Second remark concerns the scalability of $k_{kin}$. In Fig.\ref{Fig:e1e2}(a) we see that
value of kinetic wave vector is compressed for the lower subband and expanded for the second
one with respect to canonical wave vector value.
And last, the two lowest subbands become degenerated for much larger Fermi energies what
is showed in Fig.\ref{Fig:e1e2}(b). It means that electrons in both subbands
move with the same group velocity $v_{gr}=\hbar k_{kin}/m^{*}$.
This fact can be easily explained if energy dispersion relation
given by Eq.\ref{Eq:k0} will be expressed as function of $k_{kin}$ instead of k.
For this purpose, we first calculate an expectation value of kinetic wave vector for p-th subband:
\begin{equation}
 \langle k_{kin}^{(p)}\rangle=k-\sum_{l=1}^{M}\sum_{m=1}^{M}c^{(p)}_{l,1}c^{(p)}_{m,1}
 \frac{m^{*}\omega_{y}}{\hbar}Z_{l,m}^{(1)}
\end{equation}
where coefficents $c^{(p)}_{l,1}$ and $c^{(p)}_{m,1}$ are components of two-element 
effective Hamiltonian ($p=1,2$) given by  Eq.\ref{Eq:hi}. Using the components of two orthogonal
eigenvectors, after some algebra we obtain a formula for the kinetic wave vectors:
\begin{equation}
 \langle k_{kin} \rangle=k\left(1\mp \frac{m^{*}}{2\hbar^2}
 \frac{A_{2}^{2}}{\sqrt{A_{1}^{2}+A_{2}^{2}k^2}}
\right)
\end{equation}
where $(-)$ stands for first (lower) subband while $(+)$ for the second (upper) one.
Assuming very large k value we may expand expression with square root leaving only the first term
and then rearrange equation to get k:
$k\approx k_{kin}\pm \frac{m^{*}}{2\hbar^2}|A_{2}|$. Substitution this approximate expression for k
into Eq.\ref{Eq:e21} gives:
\begin{equation}
 E\left(k_{kin}^{(p)}\right)=\frac{\hbar^2}{2m^{*}}\left( k^{(p)}_{kin}
\right)^2+A_{3}-\frac{m^{*}}{8\hbar^2}A_{2}^{2}
\label{Eq:ekf}
\end{equation}
Now, it is easily to notice that if the kinetic wave vectors for the first and second subbands
have the same value then these subbands are degenerated [Fig.\ref{Fig:e1e2}(b)].

\subsection{Inter-layer subbands mixing for a non-symmetric confining potential}
\label{Sec:mbz0}
\begin{figure}[htbp!]
\hbox{
	\epsfxsize=80mm
        \epsfbox[0 397 271 842] {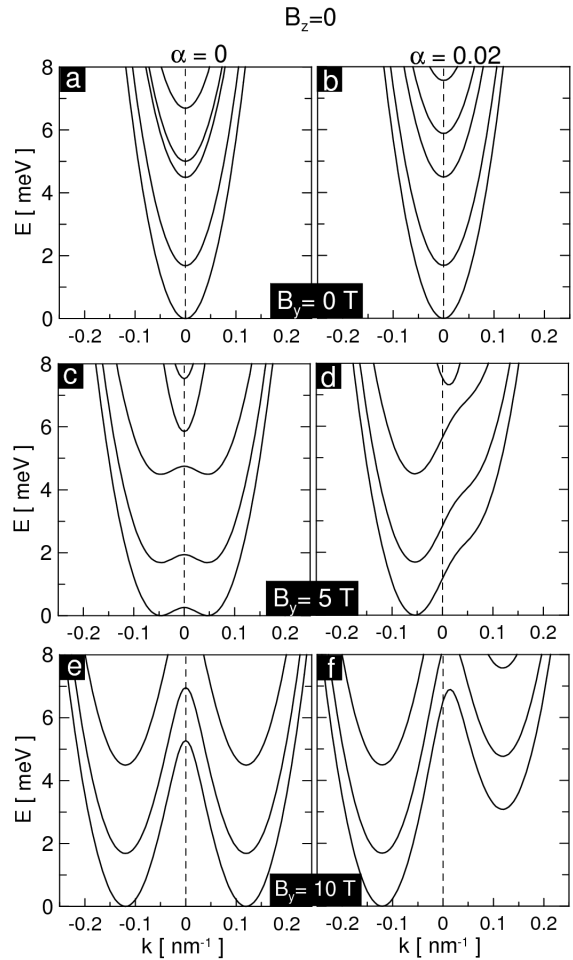}
        \hfill}
\caption{(Color online)
Energy dispersion relation for $\alpha=0$ ($\Delta E_{21}^{(z)}=5\ \textrm{meV}$, left column) and
$\alpha=0.02$ ($\Delta E_{21}^{(z)}=5.88\ \textrm{meV}$, right column).
All energies are given with respect to a bottom of the lowest subband.}
\label{Fig:e5}
\end{figure}
In this section we consider an effect of pure inter-layer subbands mixing ($B_{z}=0$) on the energy
dispersion relation for $\alpha>0$ and on the conductance of a bi-layer nanowire.
In Fig.\ref{Fig:e5} we have plotted the low energy spectra
calculated within our model for $\alpha=0$ and $\alpha=0.02$.
If there is no in-plane magnetic field [first row in Fig.\ref{Fig:e5}], the vertical
eigenmodes are not mixed and the symmetry of energy dispersion i.e. $E(k)=E(-k)$ is kept
independently of $\alpha's$ value.
For $\alpha=0.02$ the energy branches of subsequent subbands are only
shifted upwards on energy scale in comparison to the case with $\alpha=0$. When value of $B_{y}$ is
increased to $5\textrm{T}$ [second row in Fig.\ref{Fig:e5}], then for $\alpha=0$ the negative
energy dispersion relation appears ($k_{m}>0$) in the three lowest energy subbands which
correspond to the ground, the first and the second excited states for y direction.
However, the difference between the maximum and minimum of energy for each of these subbands is
very small since it equals only $0.25\textrm{ meV}$. If potential $V(z)$ becomes slightly
nonsymmetric for $\alpha=0.02$, it destroys the symmetry of subbands too. In such case,  the upper
quantum well is wider than the lower one what leads to a larger energy separation between the basis
states
$f_{1}(z)$ and $f_{2}(z)$ which for $\alpha=0.02$ is $\Delta E_{21}^{(z)}=5.88\textrm{meV}$.
Although, this growth is not large, it is sufficient to suppress
the negative energy dispersion relation for the moderate value of an in-plane magnetic field
($B_{y}=5 \textrm{T}$) since it is too small to overcome $B_{y}^{min}$ defined in
Eq.\ref{Eq:bmin}.
In this case, the electrons which have $k>0$ and are localized in the
lower layer due to an action of the magnetic force, have higher energies than those with $k<0$
[see localization of electron densities in Fig.\ref{Fig:vz}(b)] which move in a wider upper well.
For this reason, the right parts of the energy spectra in Figs.\ref{Fig:e5}(d) and (f) are shifted
upwards with respect to their left parts.
Stronger magnetic field ($\textrm{B}_{y}=10\textrm{ T}$) enhances the
negative energy dispersion in relation to $\alpha=0$ case.
In Fig.\ref{Fig:e5}(e) we may notice that the difference between maximum and minimum of energy
equals 5 meV for three lowest subbands and is much larger when compared to fraction of meV we have
got for $B_{y}=5\textrm{ T}$. Mixing of vertical subbands is now so strong that
the negative dispersion relation of energy is reconstructed also for $\alpha=0.02$ [cf.
Figs.\ref{Fig:e5}(d) and (f)].
\begin{figure}[htbp!]
\hbox{
	\epsfxsize=80mm
        \epsfbox[10 195 360 750] {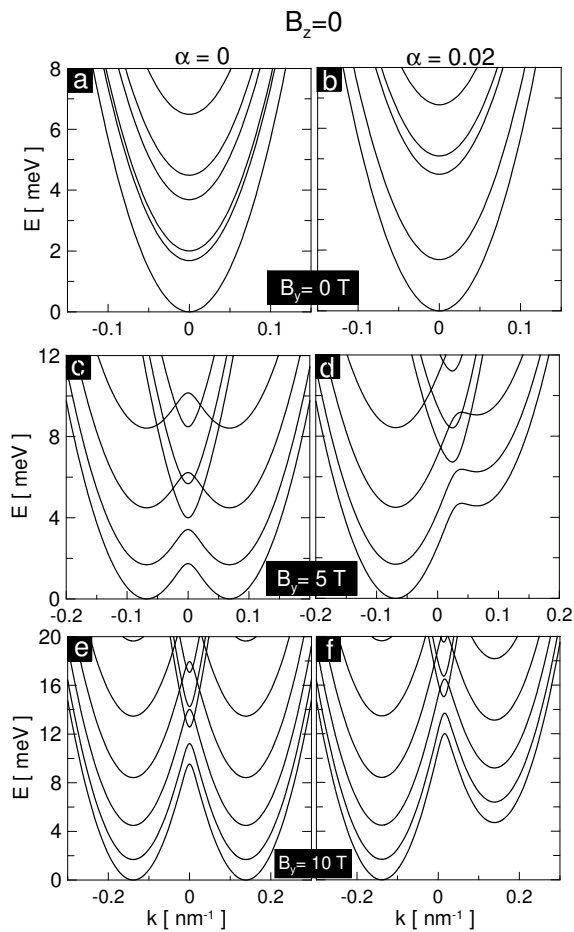}
        \hfill}
\caption{(Color online) Energy dispersion relation for $\Delta E_{21}^{(z)}=2\ \textrm{meV}$ (left
column) and $\Delta E_{21}^{(z)}=5.09\ \textrm{meV}$ (right column) and
$B_{z}=0$. Values of $\alpha$ are given on top of each columns. All
energies are given with respect to a bottom of the lowest subband.}
\label{Fig:e2}
\end{figure}

Since mixing of the vertical modes and an appearance of additional lateral energy minima in
magnetosubbands depend on $\Delta E_{21}^{(z)}$, we have repeated calculations for its smaller
value. The left column in Fig.\ref{Fig:e2} display the energy spectra for $\Delta
E_{21}^{(z)}=2\textrm{ meV}$ and $\alpha=0$. Again, in the absence of an in-plane magnetic field
component ($B_{y}=0$), subbands have parabolic shape but with lower energy spacings between
neighbours. Now however, in contrary to the previous case, a moderate in-plane magnetic field
($B_{y}=5\textrm{T}$) effectively mixes the vertical eigenmodes what leads to a formation of two
equally deep minima for $\alpha=0$ and one deep (the left one) and one shallow (the right one) for
$\alpha=0.02$.
Note also that the subbands with negative energy dispersion relation which are lying higher on
energy scale, cross with subbands of parabolic shape. These
parabolic branches are formed when an electron occupies first excited state in vertical direction
[in Eq.\ref{Eq:psipk} value of m is then reduced to  $\textrm{m}=M=2$]. These subbands are not
mixed by magnetic field and have therefore different parity than those which were already mixed.
These energy crossings survive also for $\alpha=0.02$ [cf. Figs.\ref{Fig:e2}(d) and
\ref{Fig:e2}(f)].
Let us notice here, that such crossings are absent in the low energy spectra presented in
Fig.\ref{Fig:e5}(c) due to the larger value of $\Delta E_{21}^{(z)}$ as well as  due to a limited
range of the energy scale in Figs. \ref{Fig:e5}(e) and \ref{Fig:e5}(f).
For stronger magnetic field ($B_{y}=10\textrm{ T}$) the energy minima
for $\alpha=0$ become more than five times deeper than for moderate field and almost two times
deeper than those for $\Delta E_{21}^{(z)}=5\textrm{ meV}$ [cf. Figs. \ref{Fig:e2}(e) and
\ref{Fig:e5}(e)].  
\begin{figure}[htbp!]
\hbox{
	\epsfxsize=80mm
        \epsfbox[50 255 530 485] {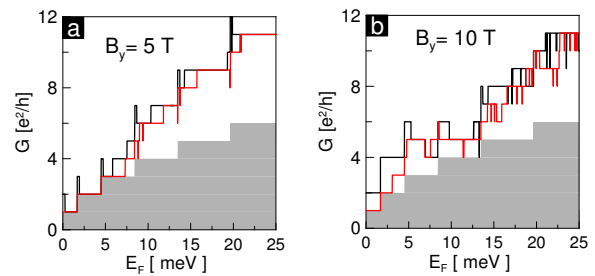}
        \hfill}
\caption{(Color online) Spin-up conductance of a nanowire for  a) $B_{y}=5\ \textrm{T}$,
b) $B_{y}=10\ \textrm{T}$  and $B_{z}=0$. Results were obtained
for a single layer wire [$M=1$, grey region] and for a bilayer wire [$M=2$,
black and red colors]. Black colour marks conductance for $\alpha=0$ ($\Delta
E_{21}^{(z)}=5\textrm{ meV}$) while red for
$\alpha=0.02$ ($\Delta E_{21}^{(z)}=5.88\textrm{ meV}$).}
\label{Fig:g5}
\end{figure}
\newline
In Fig.\ref{Fig:g5} we have plotted the conductance of a bi-layer wire
for spin-up electrons and that of a single-layer wire for
comparison. Conductance has been calculated for temperature $T=0$ by counting the crossings
between the horizontal line at Fermi energy level with those parts of subbands 
which have $k_{kin}>0$.
Results obtained for a single-layer wire [$M=1$] show standard,
well-known  step-like raising function (grey colour) with hights of steps equal to the conductance
unit
$G_{0}=e^2/h$ [the spin degeneracy is lifted]. On the other hand, the conductance for a bi-layer
wire [$M=2$, red colour stands for $\alpha=0$ while the black for $\alpha=0.02$] still
exhibits a step-like character but now two additional features appear: i) the heights of the steps
may be equal to $G_{0}$ or $2G_{0}$ and, ii) the value of conductance drops by $G_{0}$ when the
Fermi energy exceeds the height of the central maximum in a particular subband.
Due to a stronger coupling of the vertical modes for $B_{y}=10\textrm{ T}$, the rising and the
falling steps become better separated than for $B_{y}=5 \textrm{T}$. This is easily noticeable if
we compare e.g. the changes in conductance for $E_{F}<10\textrm{ meV}$ in Figs.\ref{Fig:g5} (a) and
(b).
When a confining potential looses spatial symmetry ($\alpha=0.02$), the conductance of bi-layer
nanosystem is generally lower than for $\alpha=0$ and approaches in some points a lower limit
established by the value of conductance of a single-layer wire. In both cases, even though
the conductance is significantly larger for bi-layer wire than for a single-layer case, their ratio
very rarely reaches its upper limit which equals 2.

\subsection{Mixing of the vertical and the transverse modes in tilted magnetic field}
\label{Sec:bz1}
If we account the vertical component of magnetic field ($B_{z}>0$) in our
considerations, then the term $Y_{i,j}^{(2)}$ appearing in the diagonal part of effective
Hamiltonian [Eq.\ref{Eq:hef}] becomes responsible for mixing of the transverse eigenstates
$\varphi_{i}(y)$  while the transverse and vertical eigenmodes can be simultaneously
mixed by matrix elements $H_{3}^{eff}$ defined in Eq.\ref{Eq:hef3} if $\omega_{y}\omega_{z}\ne
0$. The latter term contribute to the off-diagonal elements in Eq.\ref{Eq:hefb} for $\alpha=0$ due
to non-zero value of $Z_{1,2}^{(1)}$  or simultaneously to the diagonal and off-diagonal submatrices
in Eq.\ref{Eq:hefb} for $\alpha>0$.
\begin{figure}[htbp!]
\hbox{
	\epsfxsize=80mm
        \epsfbox[330 480 560 740 ] {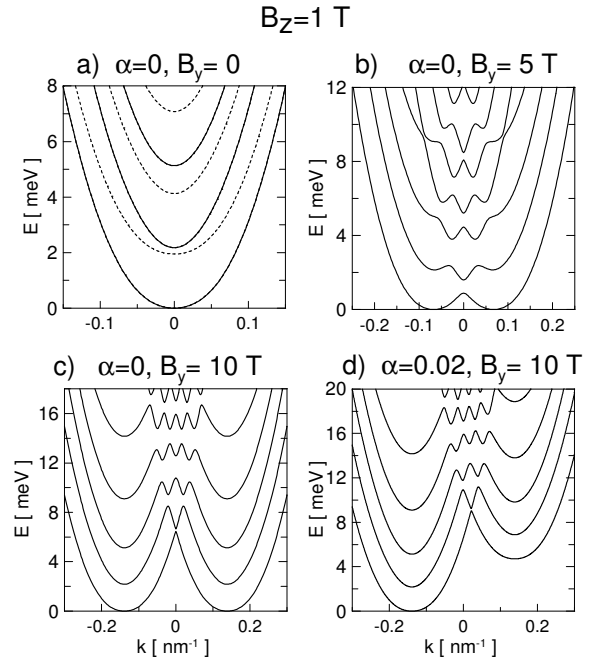}
        \hfill}
\caption{Energy dispersion relation for $\Delta E_{21}^{(z)}=2\ \textrm{meV}$ and
$B_{z}=1\ \textrm{T}$ for different values of $B_{y}$. Figures (a-c) show the results
for $\alpha=0$ while figure (d) for $\alpha=0.02$. In (a) solid lines mark these subbands in which
an electron occupies the ground state in growth direction [$f_{1}(z)$] while the dashed ones stand
for the first exicited state [$f_{2}(z)$].
All energies are given with respect to a bottom of the lowest subband.}
\label{Fig:e2bz}
\end{figure}

Magnetosubbands for $\Delta E_{21}^{(z)}=2\textrm{meV}$ and $B_{z}=1\textrm{ T}$ are plotted in
Figs.\ref{Fig:e2bz}(a-c) for $\alpha=0$ and in Fig.\ref{Fig:e2bz}(d) for $\alpha=0.02$. For
$B_{y}=0$, subbands have parabolic-like shapes and do not cross each other.
Since $\omega_{y}=0$, all the off-diagonal elements [see Eqs.\ref{Eq:hef1}-\ref{Eq:hef3}] in
effective Hamiltonian [Eq.\ref{Eq:hefb}] vanish and only the diagonal
elements given by Eq.\ref{Eq:hef0} survive but they cannot mix the vertical modes. Therefore,
the subbands corresponding to vertical excitation of an electron [dashed lines
in Fig.\ref{Fig:e2bz}(a)] are simply the replicas of those subbands in which the electron 
occupies the ground state in vertical direction. They are only shifted upwards on energy scale by
$\Delta E_{21}^{(z)}$. The wave functions of these subbands have different symmetries with respect
to reflection $z\to -z$ and for this reason may mutually intersect in higher part of energy
spectrum for $B_{y}=0$ [not shown in Fig.\ref{Fig:e2bz}(a)].
If both $B_{y}$ and $B_{z}$ have non-zero values, the off-diagonal elements in effective
Hamiltonian which are defined by Eq.\ref{Eq:hef3} may mix the vertical and transverse modes
since the products of terms $Y_{i,j}^{(1)}$ and $Z_{l,m}^{(1)}$ have non-zero values.
In Fig.\ref{Fig:e2bz}(b) we see that for $B_{y}=5\textrm{ T}$  there are two deep lateral minima in
the lowest subband similarily as in the case for $B_{y}>0$ and $B_{z}=0$ [cf. Figs. \ref{Fig:e5}(c)
and
\ref{Fig:e2}(c)].
However, for $B_{z}=1\textrm{T}$, due to mixing of the transverse modes, the 
subbands lying higher on energy scale are not the replicas of the lowest one any longer. In vicinty
of $k=0$,
crossings are replaced by avoided crossings at positions where they maximally mix with their
neighbours what in consequence  opens additional pseudogaps in the energy spectrum.
The widths of antycrossings are dependent on $B_{y}$ strength.
If value of $B_{y}$ is increased, then the spatial localizations of subbands in the lower and
upper layers are enhanced. It diminishes the overlaps between transverse modes
belonging to different layers what in consequence decreases the widths of anticrossings [cf. Figs.
\ref{Fig:e2bz}(b) and \ref{Fig:e2bz}(c)].
Generally, the pattern of these avoiding crossings does not change much when th upper and lower
quantum wells are slightly different [$\alpha=0.02$] besides an accuracy of  additional
positive slope in energy [cf. Figs. \ref{Fig:e2bz}(c) and \ref{Fig:e2bz}(d)].
Even though, for $B_{y}=5 \textrm{ T}$  the vertical modes shall be effectively mixed, the two deep
lateral minima are present only in two lowest subbands. In upper subbands they are replaced by
bending points due to a comparable mixing of the vertical and transverse modes. However, when
the transverse component of magnetic field becomes two times stronger ($B_{y}=10 \textrm{ T}$), the
coupling
of vertical modes dominates in the system and these minima are again visible in the
lowest subbands [cf. Figs. \ref{Fig:e2bz}(c) and \ref{Fig:e2bz}(d)]. In addition,
the minima in question are localized at points $\pm k_{m}$ that is exactly as in the lowest
subband [Fig.\ref{Fig:e2bz}(c)].
\begin{figure}[htbp!]
\hbox{
	\epsfxsize=60mm
        \epsfbox[135 188 510 552] {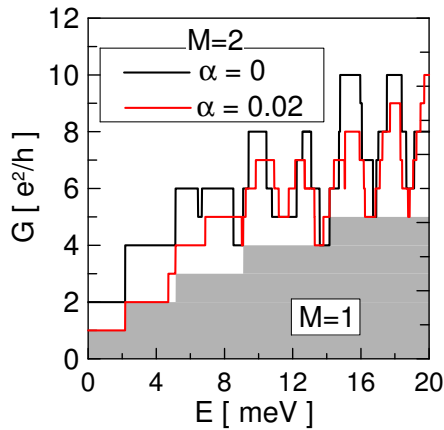}
        \hfill}
\caption{(Color online) Spin-up conductance of a single layer ($M=1$) and bi-layer wire ($M=2$)
calculated  for $B_{y}=10\ \textrm {T}$ and $B_{z}=1\ \textrm {T}$.}
\label{Fig:g2bz}
\end{figure}

An appearance of additional pseudogaps in energy spectrum significantly modifies the conductane of a
bi-layer wire. An example dependence of wire's conductance on Fermi energy is shown in
Fig.\ref{Fig:g2bz} for $B_{z}=1\textrm{ T}$ and $B_{y}=10 \textrm{ T}$. For symmetric confinement
($\alpha=0$) the conductance raises by $2G_{0}$ when the energy exceeds the bottoms
of the first three subbands [i.e. for $E=0 \textrm{ meV}$, $E=2.18 \textrm{ meV}$ and $E=5.13
\textrm{ meV}$ in Fig.\ref{Fig:g2bz}] which are determined by their lateral energy minima and
it drops by $G_{0}$ for an energy exceeding next the central maximum in particular subband.
When the energy is increased and passes through the pseudogap and then through the local energy
minima localized in vicinity of $k=0$, the value of conductance again jumps. Now however, the
heights of G steps depend on the number of local minima belonging to a particular subband
which equals the subband's index plus one. For example, in Fig.\ref{Fig:g2bz} we see that the
conductance grows by $4G_{0}$ when energy increases from $9.1\textrm{ meV}$ to $9.4\textrm{ meV}$
and even by $6G_{0}$ when it is changed between $14.2\textrm{ meV}$ and $14.8\textrm{ meV}$.
On the other hand, it may also significantly drop if an electron's energy is shifted through the set
of local maxima localized in proximity of $k=0$ because it enters a pseudogap region then.
We notice such a large drop, being equal to $5G_{0}$, when Fermi energy is increased
from $15.95\textrm{ meV}$ to $16.7\textrm{ meV}$. Generally, the conductance steps for $\alpha=0.02$
are lower than for $\alpha=0$ mainly due to a considerable shift of energy levels ladder in the
narrower layer. This effectively suppresses the mixing of vertical modes because the value of
$\Delta E_{21}^{(z)}$ significantly grows then.  The conductance of a single  layer wire
constitutes a lower bound for conductance of bi-layer wire. Both can be equal only if Fermi energy
is localized within a  pseudogap. Such case is visible in Fig.\ref{Fig:g2bz} for both
values of $\alpha$ and energies: $13.6\textrm{ meV}$ and $16.7\textrm{ meV}$.

\section{Spin polarization of conductance in bi-layer nanowire}
\label{Sec:pol}
\begin{figure*}[htbp!]
\hbox{
	\epsfxsize=120mm
       \rotatebox{-90}{{ \epsfbox[60 60 550 600] {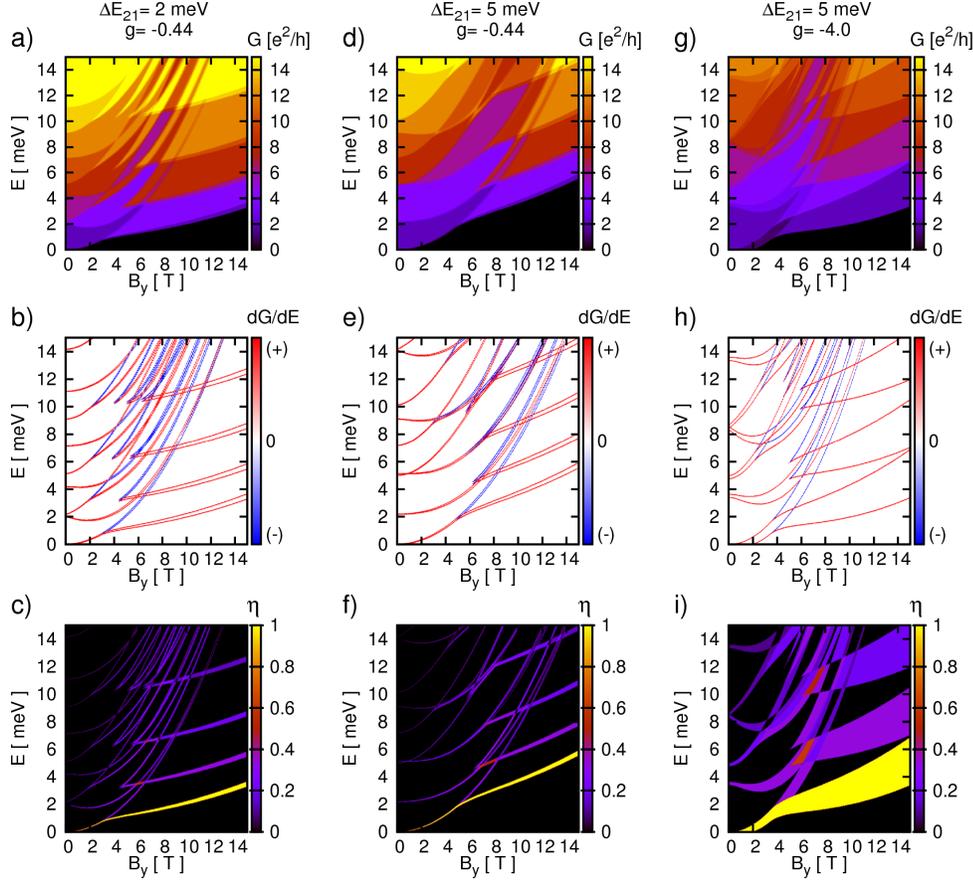}}}
        \hfill}
\caption{(Color online) Conductance (1-st row), its derivative $dG/dE$ (2-nd row) and
spin polarization of conductance (3-rd row) for bi-layer quantum wire. First and second columns
presents the results for GaAs while third column for InGaAs. The energy difference $\Delta
E_{21}^{(z)}$ is displayed on top of each columns. Value of vertical component of magnetic field
equals $B_{z}=1\textrm{ T}$. All energies are shifted down so as to the lowest subband has zero
energy for $B_{y}=0$. Therefore, in order to get the Fermi energy, an energy of the lowest subband
must be subtracted before.}
\label{Fig:t0}
\end{figure*}
\begin{figure}[htbp!]
\hbox{
	\epsfxsize=70mm
       \rotatebox{-90}{{ \epsfbox[1 350 410 840 ] {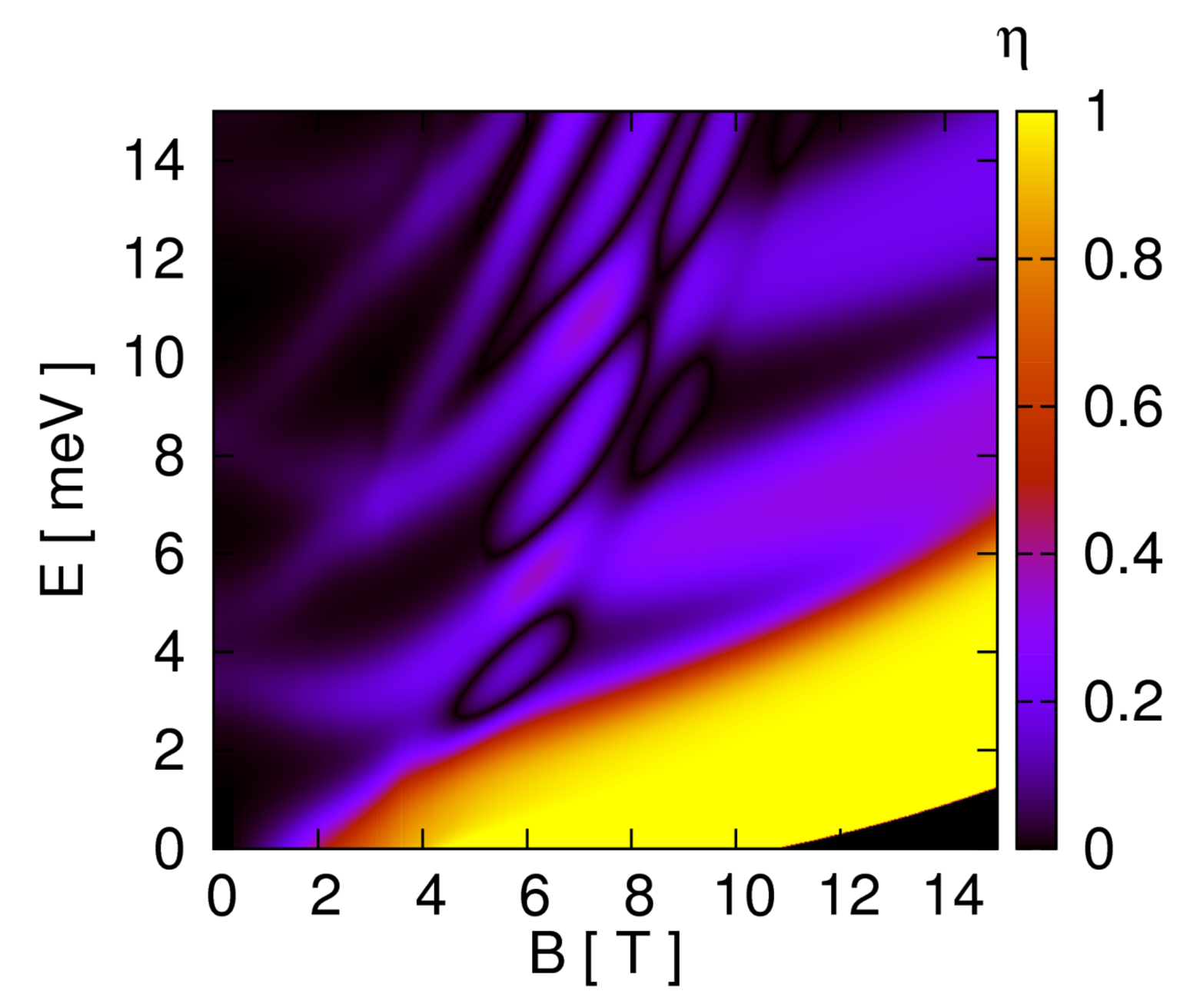}}}
        \hfill}
\caption{(Color online) Spin polarization of conductance for $\textrm{InGaAs}$ bi-layer
quantum wire for $\textrm{T}=4.2\textrm{ K}$. Other parameters are the same as for
Fig.\ref{Fig:t0}(i): $g=-4.0$ and $\Delta E_{21}^{(z)}=5\textrm{ meV}$.}
\label{Fig:t4}
\end{figure}
In previous sections we have discussed the process of formation of pseudogaps in
electron's energy spectra and its influence on the wire's conductance but have neglected the spin
Zeeman
effect contribution to energy. An interaction of electron's spin with strong magnetic field, what
is the case considered here, splits the spin-down and spin-up  subbands by  $\Delta
E_{Z}=g\mu_{B}B$. For this reason, the conductance of bi-layer wire may be partly spin polarized
\cite{qpc_imp} i.e. $\eta =(G_{\uparrow}-G_{\downarrow})/(G_{\uparrow}+G_{\downarrow})>0$ and this
polarization shall be dependent not only on a number of active spin-up and spin-down subbands as
it is in the case of  a single-layer wire but also on that whether the Fermi energy is pinned
within the pseudogap or not. In the later case one may expect a larger value of $\eta$.
 In Fig.\ref{Fig:t0} we have plotted the conductance, its derivative with respect to energy
($dG/dE$) and value of spin polarization of conductance ($\eta$) for bi-layer wire in 
function of $B_{y}$ and energy for $B_{z}=1\textrm{ T}$ and $T=0\textrm{ K}$.
These outcomes were obtained for the wire made of GaAs [first and second columns for $\Delta
E_{21}^{(z)}=2.0,\ 5.0\textrm{ meV}$ and $g=-0.44$] and of InGaAs\cite{ingaas2} [third column for
$\Delta E_{21}^{(z)}=5\textrm{ meV} $ and $g=-4.0$].

In three figures \ref{Fig:t0}(a,d,g) which show the conductance, we may notice two
characteristic regions lying above and under the anti-diagonal. In first region (above
the anti-diagonal), the changes of conductance values are frequent and can be increased as well as
decreased when electron's energy grows. The second
characteristic region appears rather for strong magnetc field (under anti-diagonal). It has more
regular pattern resembling very much that of a single layer quantum wire as the value of 
conductance increases by $G_{0}$ when subsequent subband becomes active.
Besides the conductance, also a transconductance is very often measured in experiment as it
directly reveals the dynamical properties of nanosystem being sensitive to the variations of
voltages applied to metallic gates\cite{abstreiter1} used e.g. to tune the Fermi energy in
the wire. Figures \ref{Fig:t0}(b,e,h) show dependence of similar quantity i.e.
$\textrm{dG}/\textrm{dE}$ on $\textrm{B}_{y}$ and energy values. For GaAs, that has low value of g
factor, subsequent spin-up and spin-down subbands are gathered in pairs even for strong $B_{y}$ due
to a small energy splitting [Figs.\ref{Fig:t0}(b),(e)]. However,
because of small Zeeman energy splitting, conductance becomes partly polarized only for a very
narrow energy stripes what show Figs.\ref{Fig:t0}(c) and \ref{Fig:t0}(f). Despite this fact, we
have found that even for GaAs wire, conductance can to a large extent be spin polarized. For
example, for
$\Delta E_{21}^{(z)}=2\textrm{meV}$ [Fig.\ref{Fig:t0}(c)] polarization may reach $60\%$ and $50\%$
for pairs of parameters: $B_{y}=5.94\textrm{ T}$, $E=3.54\textrm{ meV}$ and $B_{y}=6.69\textrm{ T}$,
$E=6.63\textrm{ meV}$, respectively. Regions with similar values of polarization we have also found
in Fig.\ref{Fig:t0}(f) for $\Delta E_{21}^{(z)}=5\textrm{ meV}$.
\newline
If a bi-layer nanowire is made of InGaAs which has much larger g factor than
GaAs, then an energy splitting due to spin Zeeman effect becomes even comparible with an energy
difference between the bottoms of two neighbouring subbands [see the right part of
Fig.\ref{Fig:t0}(h)]. In such case, subbands with negative energy dispersion relation (blue curves)
are shifted significantly on energy scale even for moderate magnetic field [e.g. $B_{y}\approx
4-8\textrm{ T}$ in Fig.\ref{Fig:t0}(h)].
For this reason, the regions of partly spin polarized conductance appearing for a wire
with low g factor in form of narrow stripes, now become much wider [see two distinct reddish stripes
appearing near the
central part of Fig.\ref{Fig:t0}(i), which mark $60\%$ and $50\%$ conductance polarization,
respectively]. 
This  example show the advantage of bilayer quantum wire over e.g. Y-shaped
nanostructures\cite{cummings,wojcik} in preparing partially spin polarized current. Bi-layer
nanowire enables one to get partially spin polarized current not only for the
lowest subband but also for those lying higher on energy scale  giving thus larger conductances and
currents. Drawback of this solution is however that, it still requires a strong magnetic
field to work.

We have repeated  calculations for spin polarization of conductance for InGaAs wire for
temperature $T=4.2\textrm{ K}$. Results are displayed in Fig.\ref{Fig:t4}. As expected, the
temperature smearing of subbands makes the originally moderately spin polarized regions  smaller and
additionally, it lowers their polarization to about $30\%$. This unvafourable effect  can however
be limited if semiconductor materials with much larger g factor like e.g. InSb \cite{insb} or
InAs\cite{inas} are to be used for the nanowire fabrication process.

\section{Conclusions}
\label{Sec:con}

We have theoretically investigated an effect of the magnetic field on the inter-layer and
intra-layer
subbands mixing for two vertically aligned nanowires with a rectangular-like external
confining potential. For this purpose, a simple semi-analytical method was developed which has
enabled us to
calculate the energy subbands in dependence on electron's wave vector value. It has been showed,
that the transverse component of magnetic field, which is perpendicular to the wire's
axis but parallel to the layers, can effectively mix two lowest vertical eigenmodes what
transforms the low energy subbands' parabolas into slowly oscillating curves with two deep lateral
energy minima.
If besides the transverse component of magnetic field, the vertical one is also taken into account,
then both the vertical and transverse modes are mixed simultaneously, crossings between subbands are
replaced by avoided crossings what lifts the degeneracy between subbands in vicinity of  $k=0$ and
additional small pseudogaps appear in the energy spectrum.
A qualitatively similar behaviour of magnetosubbands were predicted for two laterally aligned
wires by Shi and Gu.\cite{shi} They showed that  only one component of magnetic field, namely the
perpendicular one, is needed for an effective mixing of all pairs of magnetosubbands which were
originally localized in the same and in different wires. Consequently, such unrestricted
hybridization of magnetosubbands gives then simultaneously both types of oscillations in energy
spectrum E(k) that is, two deep lateral minima and the small-amplitude oscillations near $k=0$. 
In contrary to a nanosystem with two laterally aligned wires, an interlayer subbands hybridization
in the vertically aligned bi-layer wire depends on the transverse component of
magnetic field only while the perpendicular one is responsible for the intralayer modes
mixing.

Irrespective of the coupling direction, if these small-amplitude oscillations appear in energy
dispersion relation E(k) then the conductance of a bi-layer wire may jump as well as drop
by a few conductance quanta provided that the confining potential has low number of
defects. When  Fermi energy is shifted
through these oscillations, value of  conductance first jumps and then falls, within a thin energy
region, even by several units of conductance. Results of our simulations show that the conductance
of bi-layer quantum wire can be spin-polarized up to $60\%$ at zero temperature.
Magnitude of spin polarization can be tuned by changing the strength of magnetic field and the value
of Fermi energy.

Although there are no direct experimental results confirming our predictions, a number of
experiments were performed for similar bi-layer nanosystems. Thomas et
al.\cite{thomas} have measured the conductance of two vertically coupled wires in dependence of the
top gate voltages for parallel and perpendicular magnetic fields. Also in work of Fischer et
al.\cite{fischer4}, the conductance of similar bi-layer nanowire with nonsymmetric vertical
confining potential was experimentally investigated. In both experiments however, the lateral
confinement was smooth whereas our predictions concern the nanowires with 
rectangular-like lateral confinement and  therefore the outcomes of calculations and the
experimental data can not be directly compared.

\section*{Acknowledgements}
The work was financed by Polish Ministry of Science and Higher Education (MNiSW)
\section*{References}
\bibliography{lit2}
\end{document}